\documentclass[11pt]{article}

\usepackage[preprint]{acl} % preprint, review, final

\usepackage{times}
\usepackage{latexsym}
\usepackage[T1]{fontenc}
\usepackage[utf8]{inputenc}
\usepackage{microtype}
\usepackage{inconsolata}
\usepackage{graphicx}
\usepackage{booktabs}
\usepackage{amsmath}
\usepackage{multirow}
\usepackage{amssymb}
\usepackage{xcolor}
\usepackage{colortbl}
\usepackage{here} % for H option
\usepackage{comment}

\title{%Enhancing Emotion Recognition in Conversations: Focusing on the True Speaker and Addressing Class Imbalance
Emotion Recognition in Multi-Speaker Conversations through Speaker Identification, Knowledge Distillation, and Hierarchical Fusion}

\author{Xiao LI, Kotaro FUNAKOSHI, and Manabu OKUMURA\\
  Institute of Science Tokyo, Tokyo, Japan\\
  \texttt{\{lixiao, funakoshi, oku\}@lr.first.iir.isct.ac.jp} \\
  }
  
\begin{document}
\maketitle

\begin{abstract}
Emotion recognition in multi-speaker conversations faces significant challenges due to speaker ambiguity and severe class imbalance.
We propose a novel architecture that addresses these issues through three key innovations: (1) a speaker identification module that leverages audio-visual synchronization to accurately identify the active speaker, (2) a knowledge distillation strategy that transfers superior textual emotion understanding to audio and visual modalities, and (3) hierarchical attention fusion with composite loss functions to handle class imbalance. Comprehensive evaluations on MELD and IEMOCAP datasets demonstrate superior performance, achieving 67.75\% and 72.44\% weighted F1 scores respectively, with particularly notable improvements on minority emotion classes.
\end{abstract}

\abovedisplayskip=3pt
\belowdisplayskip=3pt

\section{Introduction}

Human emotion recognition has emerged as one of the most fundamental and challenging problems in artificial intelligence \cite{dzedzickis2020human,saxena2020emotion,deng2021survey}, with important implications for human-computer interaction, social robotics, mental health monitoring, and conversational AI systems \cite{zhao2025review, pereira2025deep}. The development of emotionally intelligent systems has become increasingly critical as AI applications expand into domains requiring nuanced understanding of human emotional states \cite{spezialetti2020emotion,younis2024machine}.

Unlike traditional pattern recognition tasks that focus on static objects or isolated signals, emotion recognition requires understanding the complex interplay of multiple communication channels through which humans naturally express their emotional states \cite{guo2024development}. Research in psychology and cognitive science has demonstrated that emotional communication is inherently multimodal, with different modalities providing complementary and sometimes redundant information about a person's emotional state \cite{manalu2024detection}.
For instance, while spoken words may convey neutral content, facial expressions might reveal underlying frustration or sarcasm \cite{zupan2024facial}.

The computational modeling of this multimodal emotion recognition process has gained significant attention in recent years, driven by advances in deep learning and the availability of large-scale multimodal datasets, such as {IEMOCAP} \cite{busso2008iemocap} and {MELD} \cite{poria2018meld}. These resources have enabled the development of increasingly complex neural architectures that integrate textual, auditory, and visual signals through sophisticated fusion mechanisms \cite{baltruvsaitis2018multimodal}.

However, early approaches to emotion recognition focused primarily on single modalities, especially on context modeling and information extraction from the text modality, while neglecting the wealth of visual and auditory information available in video and speech. Such information includes facial keypoint trajectories, expression dynamics \cite{zhang2021coin}, intonation patterns, and prosodic variations \cite{el2011survey}, all of which are fundamental to human emotion perception in real-world interactions.

The transition toward multimodal emotion recognition has been motivated by the observation that combining multiple information sources typically leads to more robust and accurate emotion classification \cite{poria2017review}. Studies \cite{mao2020dialoguetrm, wu2019multimodal} have shown that multimodal approaches can achieve significantly better performance compared to their unimodal counterparts, particularly in challenging conditions such as low-quality audio, poor lighting, or partial occlusion. Furthermore, multimodal systems can better handle the inherent ambiguity in emotional expressions, where the same facial expression might convey different emotions depending on the context and accompanying vocal cues \cite{hinton2015distilling}.

Despite these advances, conversational emotion recognition in multi-speaker scenarios presents unique challenges that remain inadequately addressed: \textbf{(1) Speaker Disambiguation} - accurately identifying the active speaker in multi-party conversations where multiple faces may be visible; \textbf{(2) Modality Performance Gaps} - significant performance disparities between text-based and audio-visual emotion recognition methods; and \textbf{(3) Severe Class Imbalance} - real-world conversational datasets often exhibit heavy bias toward certain emotions while underrepresenting others.

We propose a comprehensive framework that addresses these challenges through three main contributions:\footnote{The code repository will be public upon acceptance.}
\textbf{(1) Speaker-Centric Processing:} We introduce LipSyncNet, an %supervised contrastive learning 
approach that identifies active speakers through audio-visual synchronization patterns, enabling precise speaker-specific emotion analysis.
\textbf{(2) Cross-Modal Knowledge Distillation:} We systematically transfer knowledge from high-performing text models to audio and visual counterparts, bridging modality performance gaps through graph-based architectures.
\textbf{(3) Hierarchical Fusion with Composite Loss:} We develop sophisticated attention mechanisms combined with composite loss functions that effectively handle severe class imbalance in conversational datasets.

\section{Related Work}

\textbf{Conversational Emotion Recognition.} Conversational emotion recognition extends beyond isolated utterance analysis to consider the contextual and interactive nature of human dialogue~\cite{poria2019emotion}. Unlike traditional emotion recognition tasks that process individual samples independently, conversational scenarios require modeling temporal dependencies, speaker interactions, and emotional dynamics throughout the conversation.

Key characteristics of conversational emotion recognition include \textbf{context dependency}, where the emotional interpretation of an utterance depends on preceding context~\cite{hazarika2018conversational}; \textbf{speaker modeling}, which requires maintaining separate emotional states and characteristics for different participants~\cite{majumder2019dialoguernn}; and \textbf{emotion dynamics}, involving the modeling of emotional transitions and contagion effects between speakers~\cite{ghosal2019dialoguegcn}.

Recent approaches have employed various neural architectures to capture conversational dynamics. Recurrent Neural Networks with attention mechanisms model temporal dependencies in conversations~\cite{hazarika2018conversational}. Hierarchical approaches use separate encoders for utterance-level and conversation-level representations~\cite{majumder2019dialoguernn}. Graph-based methods represent conversations as graphs where nodes represent utterances and edges capture relationships between them~\cite{ghosal2019dialoguegcn}.

The integration of multiple modalities in conversational settings presents additional challenges, as different modalities may have varying temporal resolutions and alignment issues~\cite{tsai2019multimodal}. State-of-the-art approaches include Multimodal Transformer architectures that use cross-modal attention to align and fuse information across modalities~\cite{rahate2022multimodal}.

\textbf{Multimodal Fusion Strategies.} Multimodal emotion recognition combines information from multiple input channels to achieve more robust and accurate emotion classification than any single modality alone~\cite{baltruvsaitis2018multimodal}. The fusion process can occur at different levels of the processing pipeline, each with distinct advantages and limitations.

\textbf{Early fusion} concatenates features from different modalities before classification~\cite{wollmer2013youtube}. This approach allows the classifier to learn joint representations and cross-modal correlations but may suffer from the curse of dimensionality and differences in feature scales across modalities~\cite{poria2017review}.

\textbf{Late fusion} trains separate classifiers for each modality and combines their outputs through techniques such as weighted voting, product rule, or learned combination functions. While this approach is more robust to modality-specific noise, it cannot capture low-level cross-modal interactions~\cite{ramakrishna2023differential}.

\textbf{Hybrid fusion} strategies attempt to combine the benefits of early and late fusion by performing fusion at intermediate levels of the processing pipeline~\cite{zadeh2017tensor}. Recent approaches have explored attention-based fusion mechanisms that dynamically weight the contribution of different modalities based on their relevance to the current sample~\cite{liang2018multimodal}.

Advanced fusion architectures include Tensor Fusion Networks~\cite{zadeh2017tensor}, which model multi-modal interactions through tensor operations, and Memory Fusion Networks~\cite{zadeh2018memory}, which use attention mechanisms to selectively retrieve relevant cross-modal information. Graph-based fusion approaches~\cite{zhang2019multimodal} represent multimodal data as graphs and use Graph Neural Networks (GNNs) to capture complex relationships between modalities.

\textbf{Speaker Diarization.} Speaker diarization is the process of determining ``who spoke when'' in an audio recording containing multiple speakers~\cite{malik2020speaks}. This task is fundamental to multi-speaker emotion recognition as it provides the necessary speaker attribution required for accurate emotion analysis.

Traditional speaker diarization systems follow a clustering approach: speech segments are first extracted through Voice Activity Detection (VAD), then speaker embeddings are computed for each segment, and finally clustering algorithms group segments belonging to the same speaker~\cite{garcia2017speaker}. Common clustering techniques include k-means, hierarchical clustering, and spectral clustering~\cite{sell2018diarization}.

Modern approaches leverage deep learning for improved speaker embeddings. i-vectors~\cite{dehak2010front} and x-vectors~\cite{snyder2018x} have become standard speaker representations, with x-vectors showing superior performance through deep neural network training. End-to-end neural diarization systems~\cite{fujita2019end} jointly optimize speaker embedding and clustering components.

SyncNet~\cite{chung2017out} pioneered the use of lip-sync information for speaker identification by learning to associate lip movements with corresponding audio signals. Subsequent work has explored various audio-visual fusion strategies and improved synchronization detection methods~\cite{afouras2018conversation}.

However, existing audio-visual diarization methods often focus on face detection and tracking without consideration of multimodal emotion recognition.
The precise temporal alignment needed for emotion analysis remains a challenge in current diarization systems~\cite{bredin2020pyannote}.

\textbf{Limitations of Existing Work.} Current approaches face three key limitations: (1) assumption of perfect speaker identification or treating it as preprocessing, leading to error propagation in multi-speaker scenarios; (2) insufficient handling of modality performance imbalances, particularly the superior performance of text over audio/visual modalities in conversational contexts; and (3) inadequate solutions for severe class imbalance in real-world conversational datasets, where minority emotions are crucial for comprehensive emotion understanding. Our work addresses these limitations through integrated speaker-centric processing, systematic cross-modal knowledge distillation, and composite loss functions specifically designed for conversational class imbalance.

\section{Method}

We propose a comprehensive multimodal conversational emotion recognition framework that addresses three key challenges: speaker identification in multi-party scenarios, cross-modal knowledge transfer, and class imbalance. Our approach consists of four main components: (1) speaker-centric processing via LipSyncNet, (2) graph-based knowledge distillation, (3) hierarchical attention fusion, and (4) composite loss functions.

\begin{figure*}[t]
  \centering
  \includegraphics
  [width=.92\linewidth]%
  {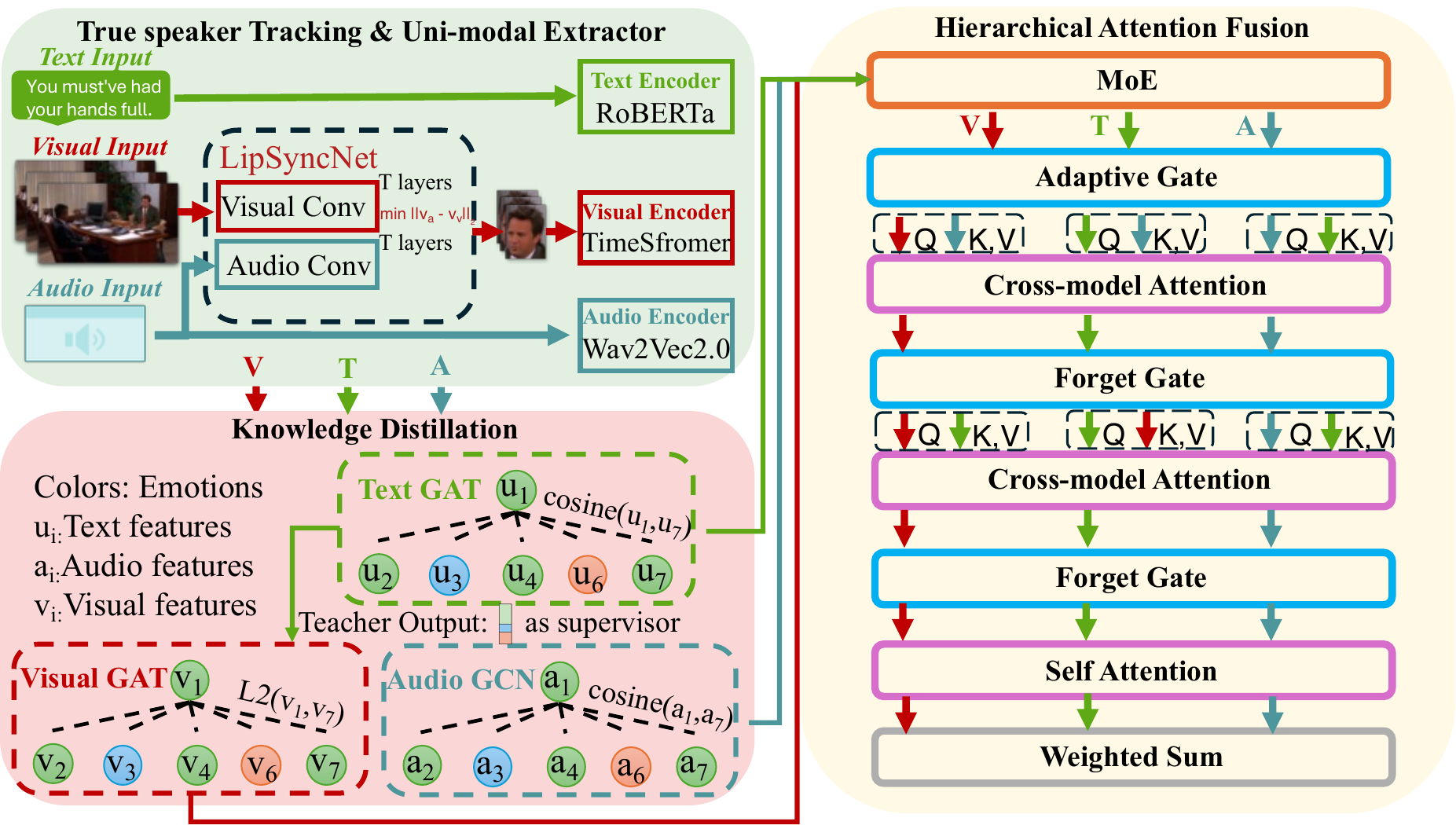}
  \caption{Overall architecture of the proposed multimodal conversational emotion recognition system.}\label{fig:system_overview}
\end{figure*}

\subsection{System Overview}

Given a dataset $\mathcal{D}{=}\{(u_i, s_i, e_i)\}_{i=1}^N$ where $u_i$ represents the $i$-th utterance, $s_i$ is the speaker identity, and $e_i$ is the emotion label, our framework extracts multimodal features $\mathbf{F}_t^{(i)} \in \mathbb{R}^{d_t}$, $\mathbf{F}_a^{(i)} \in \mathbb{R}^{d_a}$, and $\mathbf{F}_v^{(i)} \in \mathbb{R}^{d_v}$ for text, audio, and visual modalities respectively. As illustrated in Figure~\ref{fig:system_overview}, the system processes these features through modality-specific graph networks with knowledge distillation, hierarchical fusion.
The system is optimized via the following loss function:
\begin{equation}
\mathcal{L}_{total} = \mathcal{L}_{fusion} + \lambda_{dis} \mathcal{L}_{dis} + \lambda_{sync} \mathcal{L}_{sync}.
\end{equation}
The three components $\mathcal{L}_{sync}$, $\mathcal{L}_{dis}$, and $\mathcal{L}_{fusion}$ are to be explained below in this order. 

\subsection{Speaker-Centric Processing}

Multi-party conversations require accurate speaker identification to extract relevant visual cues. We propose LipSyncNet, a dual-stream architecture that learns audio-visual synchronization patterns for robust speaker filtering.

\subsubsection{Architecture and Training}

LipSyncNet employs a 3D CNN video encoder and a 2D CNN audio encoder that produce 256-dimensional L2-normalized features $\mathbf{f}_v$ and $\mathbf{f}_a$ respectively. The model is trained using a combined ranking and alignment loss:
\begin{align}
\mathcal{L}_{sync} &= \alpha_{sync}\mathcal{L}_{rank} + (1-\alpha_{sync}) \mathcal{L}_{align}, \\
\mathcal{L}_{rank} &= \max(0, m + d_{pos} - d_{neg}),\\
\mathcal{L}_{align} &= \mathbb{E}[d_{pos}],
\end{align}
where $d_{pos} = \|\mathbf{f}_v^{+} - \mathbf{f}_a\|_2$ denotes the distance between synchronized audio-visual pairs and $d_{neg} = \|\mathbf{f}_v^{-} - \mathbf{f}_a\|_2$ denotes the distance between non-synchronized pairs. $m$ is the margin. 
$\mathcal{L}_{rank}$ encourages separation between positive and negative audio-visual pairs.
$\mathcal{L}_{align}$ minimizes synchronized pair distances.

\subsubsection{Speaker Identification}

The synchronization score is defined as the negative embedding distance between a candidate face video $F_j$ and the audio $A$:
\begin{equation}
\text{Score}(F_j, A) = -\|\mathbf{f}_v^{(j)} - \mathbf{f}_a\|_2.
\end{equation}
The true active speaker is identified as the one with the highest synchronization score:
\begin{equation}
j^* = \arg\max_{j} \text{Score}(F_j, A).
\end{equation}
This ensures that subsequent visual feature extraction only relies on the true speaker rather than background participants.

\subsection{Multimodal Feature Extraction}

\subsubsection{Contextual Text Features}

Conversational emotion recognition requires understanding not only the current utterance but also the surrounding conversational context that significantly influences emotional interpretation. Unlike isolated text classification, emotions in conversations are heavily dependent on prior dialogue history, speaker relationships, and conversational flow \cite{hazarika2018conversational}.

Inspried by ~\cite{shi-huang-2023-multiemo}, we construct conversational context through a symmetric expansion strategy that captures both historical and future context around the target utterance $u_t$. This approach is motivated by the observation that emotional expressions often build upon previous exchanges while also being influenced by anticipated responses in interactive scenarios.

The context construction follows an iterative expansion process:
\begin{align}
\mathbf{C}_t &= \mathbf{L}_k \oplus [\text{SEP}] \oplus u_t\oplus [\text{SEP}] \oplus \mathbf{R}_k,
\end{align}
where $\mathbf{L}_k$ and $\mathbf{R}_k$ represent left and right context expansions within a 512-token limit imposed by RoBERTa  \cite{liu2019roberta}'s input constraints. The special separator tokens $[\text{SEP}]$ help the model distinguish the target utterance from its context.

Following \cite{yun2024telme}, we incorporate an emotion-aware prompt 
that guides the model's attention to 
emotional cues. The prompt takes the form: ``{[Speaker] feels [MASK]}'' where [Speaker] is the current speaker identity and [MASK] is the prediction target. This design encourages the model to explicitly consider speaker-specific emotional patterns within the conversational context.

The fine-tuned RoBERTa model produces contextualized hidden states 
$\mathbf{H}_{text} \in \mathbb{R}^{T_t \times 1024}$ for each token in the input sequence of length $T_t$. 
The hidden state of the final token, $\mathbf{H}_{text}[T_t] \in \mathbb{R}^{1024}$, 
is projected through a linear transformation $\mathbf{W} \in \mathbb{R}^{1024 \times 768}$ 
to obtain the utterance-level representation 
$\mathbf{F}_t^{(i)} \in \mathbb{R}^{768}$.

\subsubsection{Audio Features}

We employ Wav2Vec2.0 \cite{baevski2020wav2vec} for audio feature extraction due to its superior performance in capturing both phonetic and prosodic information essential for emotion recognition. 

The model processes raw audio waveforms and generates contextualized hidden states $\mathbf{H}_{audio} \in \mathbb{R}^{T_a \times 1024}$, where $T_a$ represents the temporal length. To obtain utterance-level representations suitable for our graph-based processing, we apply temporal average pooling:
\begin{equation}
\mathbf{F}_a^{(i)} = \frac{1}{T_a} \sum_{t=1}^{T_a} \mathbf{H}_{audio}[t, :] \in \mathbb{R}^{1024}.
\end{equation}
This pooling strategy preserves global acoustic patterns while maintaining computational efficiency for subsequent graph network processing.

\subsubsection{Visual Features}

For visual modality, we utilize TimeSformer \cite{gberta_2021_ICML} specifically designed for spatiotemporal video understanding. TimeSformer's divided attention mechanism separately models spatial and temporal relationships, making it particularly effective for capturing facial expression dynamics and micro-expressions that evolve over time in conversational contexts. This approach is superior to conventional 3D CNNs as it can model long-range temporal dependencies without the burden of dense 3D convolutions.

The model processes speaker-filtered video sequences (obtained from our LipSyncNet module) to ensure that visual features correspond to the actual speaker rather than background individuals. TimeSformer generates frame-level representations which are aggregated through temporal pooling:
\begin{equation}
\mathbf{F}_v^{(i)} = \frac{1}{T_v} \sum_{t=1}^{T_v} \mathbf{H}_{visual}[t, :] \in \mathbb{R}^{768}.
\end{equation}
The temporal pooling captures holistic facial expression patterns across the entire utterance duration, providing robust representations that are invariant to minor head movements and temporary occlusions while preserving emotional expression dynamics essential for accurate classification.

\subsection{Knowledge Distillation}

To address the performance gap between text and audio-visual modalities, we employ a teacher-student distillation framework using modality-specific graph networks. The fundamental motivation behind this approach comes from the observation that text-based emotion recognition consistently outperforms audio and visual modalities due to the explicit semantic content and rich contextual information available in linguistic expressions. 

Our graph-based approach is motivated by the inherent relational nature of conversational data, where utterances are interconnected through temporal dependencies, speaker relationships, and semantic similarities. Traditional sequence-based models often fail to capture these complex interdependencies, whereas graph neural networks can explicitly model the conversational structure and propagate information across related utterances, leading to more robust emotion recognition.

The teacher-student paradigm enables systematic knowledge transfer from the high-performing text modality to weaker audio/visual modalities. This approach not only improves individual modality performance but also enhances the overall multimodal fusion by providing more balanced and informative representations from each modality.

\subsubsection{Graph Architecture Design}

The design of modality-specific graph architectures reflects the fundamental differences in how information is structured and processed across different input channels.These architectural choices are crucial for effective knowledge transfer and optimal representation learning.

The \textbf{teacher text model} uses Graph Attention Networks (GAT) \cite{velickovic2017graph} with 4 layers and multi-head attention to capture semantic dependencies. GATs are particularly suitable for textual data due to their ability to dynamically weight the importance of different neighboring utterances based on semantic similarity and contextual relevance. The attention mechanism allows the model to focus on emotionally significant contextual information while filtering out irrelevant conversational noise.

\textbf{Student models} employ modality-appropriate architectures that respect the unique characteristics of their input domains. Graph Convolutional Networks (GCN) \cite{kipf2016semi} are used for audio processing due to homogeneous temporal patterns where adjacent audio segments typically exhibit smooth transitions and consistent acoustic properties. The spectral convolution in GCNs is well-suited for capturing these gradual variations in acoustic features across time.

For visual processing, we employ GAT to handle heterogeneous facial region relationships where different facial components (eyes, mouth, eyebrows) may have varying importance for emotion expression. The attention mechanism enables dynamic weighting of different facial regions based on their relevance to the current emotional state, allowing the model to adapt to person-specific expression patterns and cultural differences in emotional display.

\subsubsection{Distillation Loss}

The knowledge distillation process requires careful balance between preserving the student model's ability to learn from ground truth labels and incorporating the rich knowledge embedded in teacher predictions. Our composite distillation loss addresses this challenge through a weighted combination of classification and knowledge transfer objectives.

The composite distillation loss combines classification and knowledge transfer:
\begin{align}
\mathcal{L}_{dis} = &\alpha_{dis}\mathcal{L}_{ce}(f_s(\mathbf{x}_s), y) + (1-\alpha_{dis})\cdot\nonumber \\
& \mathcal{L}_{KL}\left(f_s(\mathbf{x}_s)/\tau_{dis}, f_t(\mathbf{x}_t)/\tau_{dis}\right) 
 \tau_{dis}^2,
\end{align}
 where $f_s$ and $f_t$ are student and teacher models. 

$\mathcal{L}_{ce}$ and $\mathcal{L}_{KL}$ are the standard cross-entropy loss and Kullback-Leibler divergence (see Appendix~\ref{appendix:ce-and-KL} for the definitions).

\subsection{Hierarchical Attention Fusion}

Our fusion framework addresses the fundamental challenge of integrating heterogeneous multimodal information with varying reliability, temporal alignment, and semantic granularity. The hierarchical design enables progressive information refinement through multiple processing stages, each addressing specific aspects of multimodal integration. The framework integrates multimodal features through five stages: projection, quality assessment, cross-modal attention, transformer encoding, and ensemble prediction.
The five-stage design reflects a principled approach to multimodal fusion: first standardizing feature representations, then assessing their quality, enabling cross-modal information exchange, modeling complex dependencies, and finally generating robust predictions with uncertainty estimation. This progression ensures that each stage builds upon previous refinements while adding specific capabilities essential for robust emotion recognition.

\subsubsection{Adaptive Fusion Gates}

Quality assessment and adaptive gating are critical for handling the varying reliability of different modalities under different conditions. For instance, visual features may be unreliable under poor lighting conditions, while audio features may be degraded by background noise. Our adaptive gating mechanism dynamically adjusts the contribution of each modality based on multiple quality indicators:
\begin{align}
Q^{(m)} &= \text{clamp}(\sigma(\mathbf{W}_q^{(m)} [{q}_{stats}^{(m)}; \nonumber \\
&\quad q_{entropy}^{(m)}; q_{neural}^{(m)}]), 0.1, 1.0).
\end{align}
We define three complementary indicators:  
\begin{equation}
q_{stats}^{(m)} = \frac{\sigma_{\text{inter}}^{(m)}}{\sigma_{\text{intra}}^{(m)} + \epsilon_{\text{stats}}},
\end{equation}
where $\sigma_{\text{inter}}^{(m)}$ and $\sigma_{\text{intra}}^{(m)}$ are inter-class and intra-class variances for modality $m$, 
and $\epsilon_{\text{stats}}$ is a small constant added for numerical stability. 
\begin{equation}
q_{entropy}^{(m)} = - \frac{1}{N} \sum_{i=1}^N \sum_{c=1}^C p_c^{(i,m)} \log p_c^{(i,m)},
\end{equation}
where $p_c^{(i,m)}$ is the probability of class $c$ from modality $m$ for sample $i$.  
\begin{equation}
q_{neural}^{(m)} = \sigma(\mathbf{w}_q^{(m)} \cdot \bar{\mathbf{H}}^{(m)} + b_q^{(m)}).
\end{equation}
where $\bar{\mathbf{H}}^{(m)}$ is the mean feature representation of modality $m$. 

Dynamic gates control information flow using global context as:
\begin{align}
\mathbf{H}_{gated}^{(m)} &= \mathbf{G}^{(m)} \odot \mathbf{H}^{(m)},\\
\mathbf{G}^{(m)} &= \sigma(\mathbf{W}_g^{(m)} [\mathbf{H}^{(m)}; \mathbf{C}_{global}]),\\
\mathbf{C}_{global} &= \sum_m Q^{(m)} \mathbf{H}^{(m)}/\sum_m Q^{(m)}.
\end{align}

\subsubsection{Cross-Modal Attention and Integration}

Multi-head cross-modal attention enables information exchange between modalities:
\begin{equation}
\mathbf{H}_{cross}^{(m)} = \alpha_{cross}^{(m)} \text{CrossAttn}(\mathbf{H}_{gated}^{(m)}) 
 + \beta_{cross}^{(m)} \mathbf{H}_{gated}^{(m)}.
\end{equation}
%where $\alpha_{cross}^{(m)} = 0.7$ and $\beta_{cross}^{(m)} = 0.3$. 
Before cross-modal integration, we further employ a Mixture-of-Experts (MoE) \cite{Cai_2025} layer in the fusion stage. 
These layers dynamically route modality-specific representations across multiple experts, allowing the model to capture diverse feature subspaces and provide richer inputs to the subsequent transformer encoder. 
Features are then processed through transformer encoders and hierarchical attention pooling with learnable queries to generate final predictions. 
The transformer encoders capture complex dependencies within the fused representations, while hierarchical attention pooling extracts complementary aspects of emotional information that contribute to robust classification.

\subsection{Composite Classification Loss}

Class imbalance and hard sample mining represent significant challenges in real-world conversational emotion datasets. Standard cross-entropy loss tends to be dominated by frequent classes, leading to poor performance on minority emotions that are often crucial for comprehensive emotional understanding. To address these, we propose a composite classification loss function combining polynomial focusing, label smoothing, and supervised contrastive learning.

{The polynomial loss} formulation \cite{leng2022polylosspolynomialexpansionperspective} provides an approach to hard sample mining compared to traditional focal loss. While focal loss uses exponential decay that can be too aggressive for extremely imbalanced datasets, polynomial loss offers more controlled focusing that adapts to the specific characteristics of conversational emotion data.
Our main classification loss extends cross-entropy with polynomial focusing:
\begin{equation}
\label{eq:polyloss}
\mathcal{L}_{comp} = \mathcal{L}_{ce} + \alpha_{poly} (1 - p_{y_i})^{1+\gamma_{poly}},
\end{equation}
where $p_{y_i}$ is the predicted probability for true class $y_i$, and the polynomial term provides adaptive focusing on difficult examples.

The integration of multiple loss components requires balancing to ensure that each component contributes effectively to the overall training objective without creating conflicting gradients or unstable training dynamics.
The final training objective for modality fusion combines classification, label smoothing, and supervised contrastive learning:
\begin{align}
\mathcal{L}_{fusion} &= \alpha_{comp} \mathcal{L}_{comp} + (1-\alpha_{comp}) \mathcal{L}_{smooth} \nonumber\\ 
&\quad+ \lambda_{cont} \mathcal{L}_{cont}.
\end{align}

The label smoothing loss~\cite{muller2019does} is defined as cross-entropy between the predicted distribution and a smoothed target distribution.
Supervised contrastive learning~\cite{khosla2020supervised} uses normalized feature projections to enhance discrimination between similar emotions. 
We follow the conventional use of the two objectives.
Implementation details of $\mathcal{L}_{smooth}$
and $\mathcal{L}_{cont}$
are described in Appendix~\ref{appendix:ls-and-cl} to be self-contained.

\section{Experiments}

\subsection{Datasets and Experimental Setup}

We evaluate our framework on two widely-used conversational emotion recognition benchmarks:

\textbf{MELD} \cite{poria2018meld}: A large-scale dataset extracted from TV show \textit{Friends}. The dataset exhibits severe class imbalance with neutral (47.1\%) dominating, while disgust (2.7\%) and fear (2.7\%) are severely under-represented.

\textbf{IEMOCAP} \cite{busso2008iemocap}: A carefully constructed dataset from laboratory-recorded dyadic conversations. This dataset provides higher annotation quality and more balanced class distribution compared to MELD.

Following standard protocols, we use the official train/validation/test splits for MELD and 5-fold cross-validation for IEMOCAP. 
We report weighted F1 (WF1) scores as the primary metric due to class imbalance considerations, along with class-specific F1 scores for detailed analysis.

Depending on each dataset, we compare our method against representative past works %as baselines 
\cite{ghosal2019dialoguegcn,majumder2019dialoguernn,lu-etal-2020-iterative,li2021quantum,li2021ctnet,joshi-etal-2022-cogmen,li-etal-2023-joyful,shi-huang-2023-multiemo,yun2024telme}.

\subsection{Implementation Details}

Our method is implemented using PyTorch 1.12+ and trained on a NVIDIA RTX 3090 GPU.
We use the AdamW optimizer.
Model training except for data pre-processing (i.e., KD \& Fusion parts with 3M parameters) requires 30 to 50 minutes. 
Appendix~\ref{appendix:hyperparams} shows all hyperparameter values.
We report results averaged over 5 independent runs. %with statistical significance testing. 

\subsection{Performance Analysis}

\paragraph{MELD Dataset Performance:} Table~\ref{tab:meld_detailed} presents comprehensive per-class results on MELD, revealing the effectiveness of our approach on severely imbalanced classes. On MELD, our framework achieves 67.8\% weighted F1, representing significant improvements over the strongest baseline TelME \cite{yun2024telme}. 
Our results demonstrate statistically significant improvements with p < 0.01 (Appendix \ref{appendix:stat-test}), validating the robustness of our approach. 
More importantly, we demonstrate substantial improvements on severely underrepresented emotions: disgust (29.0\% vs. 28.0\% best baseline) and fear (30.5\% vs. 24.0\%), addressing critical limitations in minority class recognition.

\begin{table}[t]
\centering
\small
\resizebox{\columnwidth}{!}{
\tabcolsep=1pt
\begin{tabular}{lccccccccc}
\toprule
\multirow{2}{*}{\textbf{Methods}} & 
\textbf{Anger} & 
\textbf{Disgust} & 
\textbf{Fear} & 
\textbf{Joy} & 
\textbf{Neut.} & 
\textbf{Sad.} & 
\textbf{Surp.} &
\multicolumn{2}{c}{\textbf{Average}} \\ 
\cmidrule(r){2-10}
& \textbf{(1109)} & \textbf{(271)} & \textbf{(268)} & \textbf{(1743)} & \textbf{(4710)} & \textbf{(683)} & \textbf{(1205)} & \textbf{Acc.} & \textbf{WF1} \\ 
\midrule
DialogueGCN & 36.6 & 6.7 & 2.9 & 44.3 & 54.7 & 21.9 & 41.8 & 41.8 & 42.5 \\
DialogueRNN & 41.5 & 1.7 & 1.3 & \textbf{73.6} & 55.9 & 24.0 & 50.7 & 41.1 & 49.4 \\
QMNN & 43.2 & 0.0 & 0.0 & 51.4 & 77.0 & 11.7 & 53.2 & 59.7 & 59.0 \\
IterativeERC & 48.9 & 19.4 & 3.3 & 56.6 & 77.5 & 23.6 & 53.7 & 61.7 & 60.1 \\
MultiEMO & 54.0 & 28.0 & 24.0 & 64.8 & 80.0 & 43.5 & 58.3 & 67.4 & 66.6 \\
TelME & 55.9 & 22.2 & 22.2 & 65.2 & \textbf{80.9} & 43.4 & 60.2 & 68.4 & 67.3 \\ 
\midrule
\textbf{Our Method}  & \textbf{57.8} & \textbf{29.0} & \textbf{30.5} & 66.7 & 79.7 & \textbf{44.6} & \textbf{61.2} & \textbf{68.8} & \textbf{67.8} \\
\bottomrule
\end{tabular}}
\caption{Performance comparison on MELD. Numbers in parentheses indicate sample counts.}
\label{tab:meld_detailed}
\end{table}

Our framework demonstrates particularly strong performance on challenging minority classes, with notable improvements in disgust (+1.0\% point) and fear (+6.5\% points) recognition compared to the best baselines. These improvements are crucial for practical applications where minority emotion detection is often most critical.

\paragraph{IEMOCAP Dataset Performance:} Table~\ref{tab:iemocap_detailed} shows our balanced performance across IEMOCAP's emotion categories, maintaining strong recognition across all emotion types. On IEMOCAP, we achieve 72.4\% weighted F1, a significant 1.2\% improvement over MultiEMO~\cite{shi-huang-2023-multiemo}, with consistent performance gains across emotion categories while maintaining balanced recognition across all classes. 
%Our results demonstrate statistically significant improvements with p < 0.01 (Appendix \ref{appendix:stat-test}), validating the robustness of our approach. 

%For additional analysis, t
\paragraph{Additional Analysis:}
The comparison result between unimodal and multimodal models is provided in Table~\ref{tab:modality_comparison} in Appendix~\ref{appendix:modality_analysis}.
Confusion matrices for the main results are shown in Appendix~\ref{appendix:confusion-matrices}

\begin{table}[t]
\centering
\small
\resizebox{\columnwidth}{!}{
\tabcolsep=1pt
\begin{tabular}{lccccccccc}
\toprule
\multirow{2}{*}{\textbf{Methods}} & 
\textbf{Happy} & 
\textbf{Sad} & 
\textbf{Neut.} & %Neutral
\textbf{Angry} & 
\textbf{Excited} & 
\textbf{Frus.} & %Frustrated
\multicolumn{2}{c}{\textbf{Average}} \\ 
\cmidrule(r){2-9}
& \textbf{(392)} & \textbf{(739)} & \textbf{(1167)} & \textbf{(711)} & \textbf{(620)} & \textbf{(1149)} & \textbf{Acc.} & \textbf{WF1} \\ 
\midrule
DialogueGCN & 42.7 & \textbf{84.5} & 63.6 & 64.1 & 63.2 & 67.0 & 65.3 & 64.2 \\
CTNet       & 51.3 & 79.9 & 65.8 & 67.3 & \textbf{78.7} & 58.9 & 68.0 & 67.5 \\
COGMEN      & 52.0 & 81.8 & 68.7 & 66.0 & 75.4 & 68.2 & 68.3 & 67.7 \\
IterativeERC & 50.2 & 77.2 & 61.3 & 61.5 & 69.2 & 60.9 & 66.2 & 64.4 \\
QMNN        & 39.7 & 68.3 & 55.3 & 62.6 & 66.7 & 62.2 & 61.5 & 59.9 \\
TelME       & 50.8 & 80.4 & 66.5 & 66.2 & 73.5 & 67.1 & 66.9 & 68.6 \\
JOYFUL      & 56.5 & 84.3 & 68.4 & 66.9 & 73.4 & 67.6 & 70.6 & 71.0 \\
MultiEMO    & 52.7 & 83.2 & 70.0 & 65.7 & 72.9 & 70.0 & 71.5 & 71.2 \\
\midrule
\textbf{Our Method}  & \textbf{57.8} & 83.0 & \textbf{71.6} & \textbf{68.3} & 72.8 & \textbf{71.9} & \textbf{71.9} & \textbf{72.4} \\
\bottomrule
\end{tabular}}
\caption{Performance comparison on IEMOCAP. Numbers in parentheses indicate sample counts.}
\label{tab:iemocap_detailed}
\end{table}

\subsection{Ablation Study}

\begin{table}[t]
\centering
\small
\begin{tabular}{lcc}
\toprule
\textbf{Configuration} & \textbf{MELD} & \textbf{IEMOCAP} \\
\midrule
% \textbf{Additive Analysis} &  &  \\
% + Knowledge Distillation & +1.4 & +1.4 \\
% + Cross-Modal Attention & +1.1 & +0.4 \\
% + Adaptive Gating & +1.0 & +0.7 \\
% + Polynomial loss & +0.9 & +0.8 \\
% \midrule
% \textbf{Subtractive Analysis} &  &  \\
w/o %Speaker-centric
Speaker Identification & -2.6 & -1.4 \\
w/o Fusion Loss & -2.0 & -1.2 \\
w/o Knowledge Distillation & -1.9 & -1.2 \\
w/o Contrastive Learning & -0.9 & -0.6 \\
\bottomrule
\end{tabular}
\caption{Ablation study results showing component contributions as $\Delta$ WF1 (points) from the performance of the proposed method shown in Table~\ref{tab:meld_detailed} and Table~\ref{tab:iemocap_detailed}.}
\label{tab:ablation}
\end{table}

Table~\ref{tab:ablation} shows the contribution of each component through systematic ablation analysis. 
%Knowledge distillation provides the largest performance gain (+1.4 F1), validating our hypothesis about cross-modal knowledge transfer effectiveness.
%``+~Polynomial Loss'' indicates that the cross-entropy term $\mathcal{L}_{ce}$ is extended to the polynomial form defined in (\ref{eq:polyloss}). 
``w/o Speaker Identification'' means that the first found face was always used.
``w/o Fusion Loss'' means that the entire fusion objective is reduced to the standard cross-entropy loss.

Speaker%-centric processing 
identification shows significant impact when removed (-1.4 to -2.6 WF1 points), highlighting the importance of accurate speaker identification in multi-party conversation scenarios. %The composite  loss function demonstrates substantial contributions to handling class imbalance, particularly evident in the severely imbalanced MELD.
Our Fusion Loss and Knowledge Distillation also show higher impact than conventional Contrastive Learning.

\section{Conclusion}

We present a comprehensive framework for multimodal conversational emotion recognition that systematically addresses speaker identification and class imbalance challenges through three key innovations. Our LipSyncNet-based speaker identification integrates audio-visual synchronization learning directly into the emotion recognition pipeline, eliminating error propagation from preprocessing steps. Cross-modal knowledge distillation successfully transfers superior textual understanding to audio and visual modalities through graph-based architectures. Hierarchical attention fusion with composite loss functions effectively handles severe class imbalance while maintaining strong overall performance.

Experimental results demonstrate state-of-the-art performance on both MELD and IEMOCAP, with particularly notable improvements on challenging minority emotions that are crucial for comprehensive emotional understanding. The framework shows practical viability with reasonable computational requirements and robust performance across different experimental conditions.

Key contributions include: (1) integrating speaker identification as a learnable component rather than preprocessing step, (2) systematic knowledge transfer across modalities using graph-based teacher-student frameworks, and (3) composite loss functions that effectively address severe class imbalance in conversational scenarios. 

\section*{Limitations}

Our evaluation focuses primarily on English conversational data from specific domains (TV shows, laboratory recordings). Cross-lingual and cross-cultural generalization requires further investigation, particularly for languages with different prosodic patterns or cultural expression norms.

The composite classification loss function introduces additional hyperparameters requiring careful tuning for different datasets and domains.

\bibliography{acl_latex}

\appendix
\clearpage
%\onecolumn

\section{Cross-entropy loss and KL divergence}
\label{appendix:ce-and-KL}.

The cross-entropy loss is defined as:
\begin{equation}
\mathcal{L}_{ce} = - \sum_{c=1}^{C} y_c \log p_c,
\end{equation}
where $C$ is the number of classes, $y_c \in \{0,1\}$ is the ground-truth one-hot label, and $p_c$ is the predicted probability. 

The Kullback–Leibler divergence is defined as:
\begin{equation}
\mathcal{L}_{KL}(p_s, p_t) = \sum_{c=1}^{C} p_t(c) \log \frac{p_t(c)}{p_s(c)}.
\end{equation}
where $p_t(c)$ and $p_s(c)$ are teacher and student probabilities after temperature scaling.

\section{Label Smoothing and Supervised Contrastive Learning}
\label{appendix:ls-and-cl}

\textbf{The label smoothing loss}~\cite{muller2019does} is defined as cross-entropy between the predicted distribution and a smoothed target distribution:
\begin{equation}
\mathcal{L}_{smooth} = - \sum_{c=1}^{C} q_c \log p_c.
\end{equation}
where $p_c$ is the predicted probability for class $c$, and $q_c$ is the smoothed target distribution. 
\begin{equation}
q_c = 
\begin{cases}
1 - \epsilon_{\text{smooth}} + \frac{\epsilon_{\text{smooth}}}{C}, & \text{if } c = y, \\
\frac{\epsilon_{\text{smooth}}}{C}, & \text{otherwise}.
\end{cases}
\end{equation}
where $y$ is the ground-truth class, $C$ is the number of classes, and $\epsilon_{\text{smooth}}$ is the smoothing ratio.
It prevents overconfidence and improves generalization by encouraging the model to be less certain about its predictions, which is particularly beneficial for minority classes where limited training data may lead to overfitting \cite{muller2019does}.

\textbf{Supervised Contrastive learning}~\cite{khosla2020supervised} uses normalized feature projections to enhance discrimination between similar emotions.
Let $\mathbf{h}_i \in \mathbb{R}^{d_f}$ denote the fused representation of the $i$-th sample before the classifier (where $d_f$ is the fusion dimension), and let $g(\cdot): \mathbb{R}^{d_f} \rightarrow \mathbb{R}^{d_z}$ be a two-layer MLP projection head.
We obtain L2-normalized projections as
\begin{equation}
\mathbf{z}_i = \frac{g(\mathbf{h}_i)}{|g(\mathbf{h}_i)|_2},
\quad
\mathbf{z}_i^{+} = \frac{g(\tilde{\mathbf{h}}_i)}{|g(\tilde{\mathbf{h}}_i)|_2},
\end{equation}
where $\tilde{\mathbf{h}}_i$ is a positive sample for $\mathbf{h}i$ obtained either from an augmentation of the same instance or from another instance with the same ground-truth label within the mini-batch.
All negatives $\mathbf{z}j$ are the projections of the remaining instances in the mini-batch with $j \neq i$.
The contrastive objective is
\begin{equation}
\mathcal{L}_{cont} = -\frac{1}{N} \sum_{i=1}^{N} \log \frac{\exp(\mathbf{z}_i^T \mathbf{z}_i^+ / \tau_{cont})}{\sum_{j \neq i} \exp(\mathbf{z}_i^T \mathbf{z}_j / \tau_{cont})},
\end{equation}
with temperature parameter $\tau{cont}$.

\section{Hyperparameter Settings}
\label{appendix:hyperparams}

\begin{table}[H]
\centering
%\small
\begin{tabular}{ll}
\toprule
\textbf{Parameter} & \textbf{Value} \\
\midrule
Batch size & 16 \\
Number of epochs & 30 \\
Dropout rate & 0.35 \\
Fusion dimension & 256 \\
Gradient clipping (maximum norm) & 1.0 \\
\midrule
Learning rate (Text) & $8 \times 10^{-5}$ \\
Learning rate (Audio) & $6 \times 10^{-5}$ \\
Learning rate (Visual) & $6 \times 10^{-5}$ \\
Learning rate (Fusion) & $4 \times 10^{-5}$ \\
Weight decay (all modules) & 0.001 \\
\midrule
$\lambda_{dis}$ & 0.3 \\
$\lambda_{sync}$ & 0.15 \\
margin $m$  & 1.5 \\
$\alpha_{sync}$ & 0.7 \\
$\alpha_{poly}$ & 1.2 \\
$\alpha_{cross}$ & 0.7 \\
$\beta_{cross}$ & 0.3 \\
$\gamma_{poly}$ & 1.2 \\
$\alpha_{comp}$ & 0.8 \\
$\lambda_{cont}$ & 0.1 \\
$\tau_{cont}$ & 0.07  \\
$\epsilon_{\text{stats}}$ & $1 \times 10^{-6}$ \\
$\epsilon_{\text{smooth}}$ & 0.1 \\
Distillation temperature $\tau_{dis}$ & 2.0 \\
Distillation balance $\alpha_{dis}$ & 0.65 \\
\bottomrule
\end{tabular}
\caption{Hyperparameter configurations used in our experiments.}
\end{table}

The balance parameter $\alpha_{dis} = 0.65$ gives slightly more weight to ground truth supervision, ensuring that student models maintain their discriminative ability while benefiting from teacher guidance. The temperature parameter $\tau_{dis} = 2.0$ softens the probability distributions, allowing the student to learn from the teacher's uncertainty patterns and confidence levels, which often contain valuable information about decision boundaries and class relationships.

\section{Modality Performance Analysis}
\label{appendix:modality_analysis}

%%% WHAT IS THIS???
% \begin{figure}[h]
%   \centering
%   \includegraphics[width=0.7\textwidth]{figures/modality_scores.eps}
%   \caption{Average unimodal performance across different modalities reported in previous studies, illustrating the stronger role of textual features and motivating cross-modal knowledge transfer.}\label{fig:modality_scores}
% \end{figure}

\subsection{Modality-Specific Performance Analysis}
Table~\ref{tab:modality_comparison} presents a comprehensive analysis of individual modality contributions and their fusion effectiveness. This analysis provides crucial insights into the relative importance of different modalities in conversational emotion recognition tasks.

\begin{table}[h]
\centering
%\small
\tabcolsep=4pt
\begin{tabular}{lcc}
\toprule
\textbf{Models} & \textbf{MELD} & \textbf{IEMOCAP} \\
\midrule
Only Visual & 37.8 & 32.5 \\
Only Visual\textsubscript{w/o LipSyncNet} & 33.4 & 30.1 \\
Only Audio & 47.3 & 48.4 \\
Only Text & 66.1 & 68.7 \\
\midrule
\textbf{Fusion Model} & \textbf{67.8} & \textbf{72.4} \\
\bottomrule
\end{tabular}
\caption{Unimodal vs. Multimodal Performance Comparison (WF1 scores).}
\label{tab:modality_comparison}
\end{table}

The results reveal several important findings about modality contributions in conversational emotion recognition:
\textbf{Text Modality Dominance:} Text emerges as the most informative modality, achieving 66.1\% and 68.7\% WF1 on MELD and IEMOCAP respectively. This demonstrates that linguistic content carries the primary emotional signals in conversational contexts, consistent with the rich semantic information available in spoken language.
\textbf{Audio Modality Contribution:} Audio modality shows moderate performance (47.3\% on MELD, 48.4\% on IEMOCAP), capturing prosodic and paralinguistic cues that complement textual information. The stronger performance on IEMOCAP suggests that laboratory-recorded conversations may preserve more nuanced acoustic features compared to TV show audio.
\textbf{Visual Modality Challenges:} Visual features show the lowest individual performance (37.8\% on MELD, 32.5\% on IEMOCAP), highlighting the inherent difficulty of emotion recognition from facial expressions alone in conversational settings. The inclusion of LipSyncNet provides a meaningful contribution (+4.4\% on MELD, +2.4\% on IEMOCAP), demonstrating the value of audio-visual synchronization features for emotion understanding.
\textbf{Fusion Benefits:} The multimodal fusion model achieves substantial improvements over the best single modality (text), with gains of +1.7\% on MELD and +3.7\% on IEMOCAP. These improvements validate our fusion strategy's effectiveness in leveraging complementary information across modalities, particularly the integration of prosodic audio cues and synchronized visual features with semantic text information.

\section{Detailed Performance Tables and Confusion Matrices}
\label{appendix:confusion-matrices}

\subsection{MELD Dataset Results}
\begin{table*}[t]
%\small
\centering
\begin{tabular}{lccccccccc}
\toprule
\multirow{2}{*}{\textbf{Methods}} & 
\textbf{Anger} & 
\textbf{Disgust} & 
\textbf{Fear} & 
\textbf{Joy} & 
\textbf{Neutral} & 
\textbf{Sadness} & 
\textbf{Surprise} &
\multicolumn{2}{c}{\textbf{Average}} \\ 
\cmidrule(r){2-10}
& \textbf{(1109)} & \textbf{(271)} & \textbf{(268)} & \textbf{(1743)} & \textbf{(4710)} & \textbf{(683)} & \textbf{(1205)} & \textbf{Acc.} & \textbf{WF1} \\ 
\midrule
DialogueGCN & 36.56 & 6.73 & 2.85 & 44.26 & 54.72 & 21.87 & 41.78 & 41.75 & 42.46 \\
DialogueRNN & 41.53 & 1.74 & 1.25 & 73.56 & 55.94 & 23.97 & 50.73 & 41.09 & 49.43 \\
QMNN & 43.17 & 0.00 & 0.00 & 51.44 & 77.00 & 11.70 & 53.18 & 59.66 & 59.03 \\
IterativeERC & 48.88 & 19.38 & 3.31 & 56.63 & 77.52 & 23.62 & 53.65 & 61.66 & 60.07 \\
MultiEMO & 54.02 & 28.00 & 24.00 & 64.79 & 80.02 & 43.45 & 58.28 & 67.39 & 66.55 \\
TelME & 55.91 & 22.22 & 22.22 & 65.24 & 80.88 & 43.41 & 60.17 & 68.35 & 67.30 \\ 
\midrule
\textbf{Our Method}  & \textbf{57.76} & \textbf{28.79} & \textbf{30.46} & 66.67 & 79.73 & \textbf{44.63} & \textbf{61.20} & \textbf{68.78} & \textbf{67.75} \\
\bottomrule
\end{tabular}
\caption{Detailed performance comparison on MELD dataset with 2 decimal places. Numbers in parentheses indicate sample counts.}
\label{tab:meld_detailed_precise}
\end{table*}

\begin{figure}[H]
  \centering
  \includegraphics[width=0.5\textwidth]{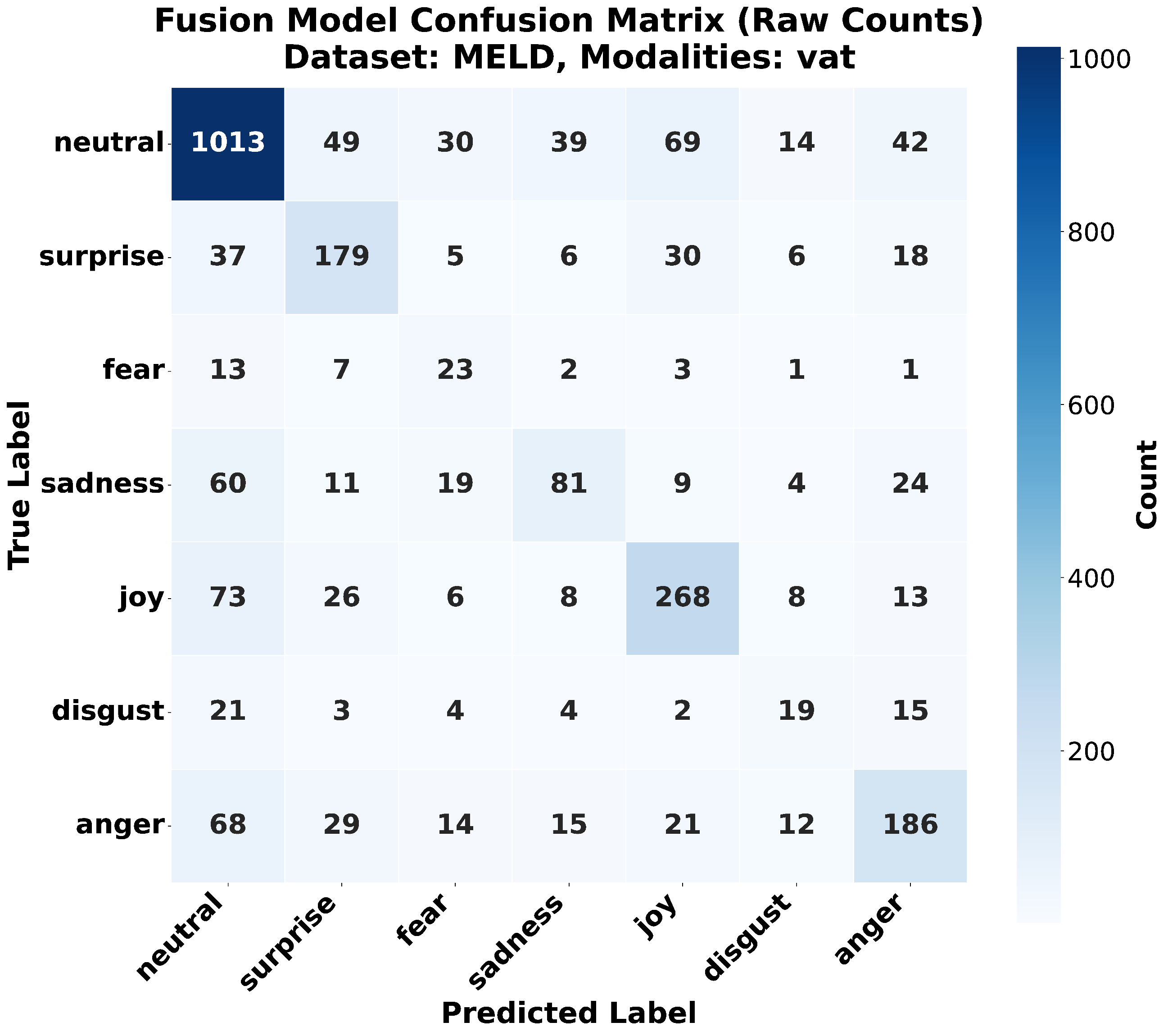}
  \caption{Confusion matrix for emotion classification on MELD dataset showing improved performance on minority emotion classes.}\label{fig:confusion_matrix_meld}
\end{figure}

\subsection{IEMOCAP Dataset Results}

\begin{table*}[t]
%\small
\centering
\begin{tabular}{lccccccccc}
\toprule
\multirow{2}{*}{\textbf{Methods}} & 
\textbf{Happy} & 
\textbf{Sad} & 
\textbf{Neutral} & %Neutral
\textbf{Angry} & 
\textbf{Excited} & 
\textbf{Frustrated} & %Frustrated
\multicolumn{2}{c}{\textbf{Average}} \\ 
\cmidrule(r){2-9}
& \textbf{(392)} & \textbf{(739)} & \textbf{(1167)} & \textbf{(711)} & \textbf{(620)} & \textbf{(1149)} & \textbf{Acc.} & \textbf{WF1} \\ 
\midrule
DialogueGCN & 42.73 & 84.52 & 63.56 & 64.13 & 63.17 & 66.95 & 65.26 & 64.18 \\
CTNet       & 51.34 & 79.93 & 65.84 & 67.26 & 78.72 & 58.89 & 68.04 & 67.53 \\
COGMEN      & 51.96 & 81.75 & 68.68 & 66.04 & 75.38 & 68.24 & 68.27 & 67.65 \\
IterativeERC & 50.17 & 77.19 & 61.31 & 61.45 & 69.23 & 60.92 & 66.21 & 64.37 \\
QMNN        & 39.71 & 68.30 & 55.29 & 62.58 & 66.71 & 62.19 & 61.52 & 59.88 \\
TelME       & 50.84 & 80.36 & 66.47 & 66.18 & 73.54 & 67.09 & 66.85 & 68.56 \\
JOYFUL      & 56.53 & 84.33 & 68.42 & 66.90 & 73.40 & 67.61 & 70.63 & 70.90 \\
MultiEMO    & 52.65 & 83.18 & 70.02 & 65.74 & 72.88 & 69.98 & 71.46 & 71.18 \\
\midrule
\textbf{Our Method}  & \textbf{57.81} & 83.04 & \textbf{71.59} & \textbf{68.25} & 72.76 & \textbf{71.93} & \textbf{71.94} & \textbf{72.44} \\
\bottomrule
\end{tabular}%}
\caption{Detailed performance comparison on IEMOCAP dataset with 2 decimal places. Numbers in parentheses indicate sample counts.}
\label{tab:iemocap_detailed_precise}
\end{table*}

\begin{figure}[H]
  \centering
  \includegraphics[width=0.5\textwidth]{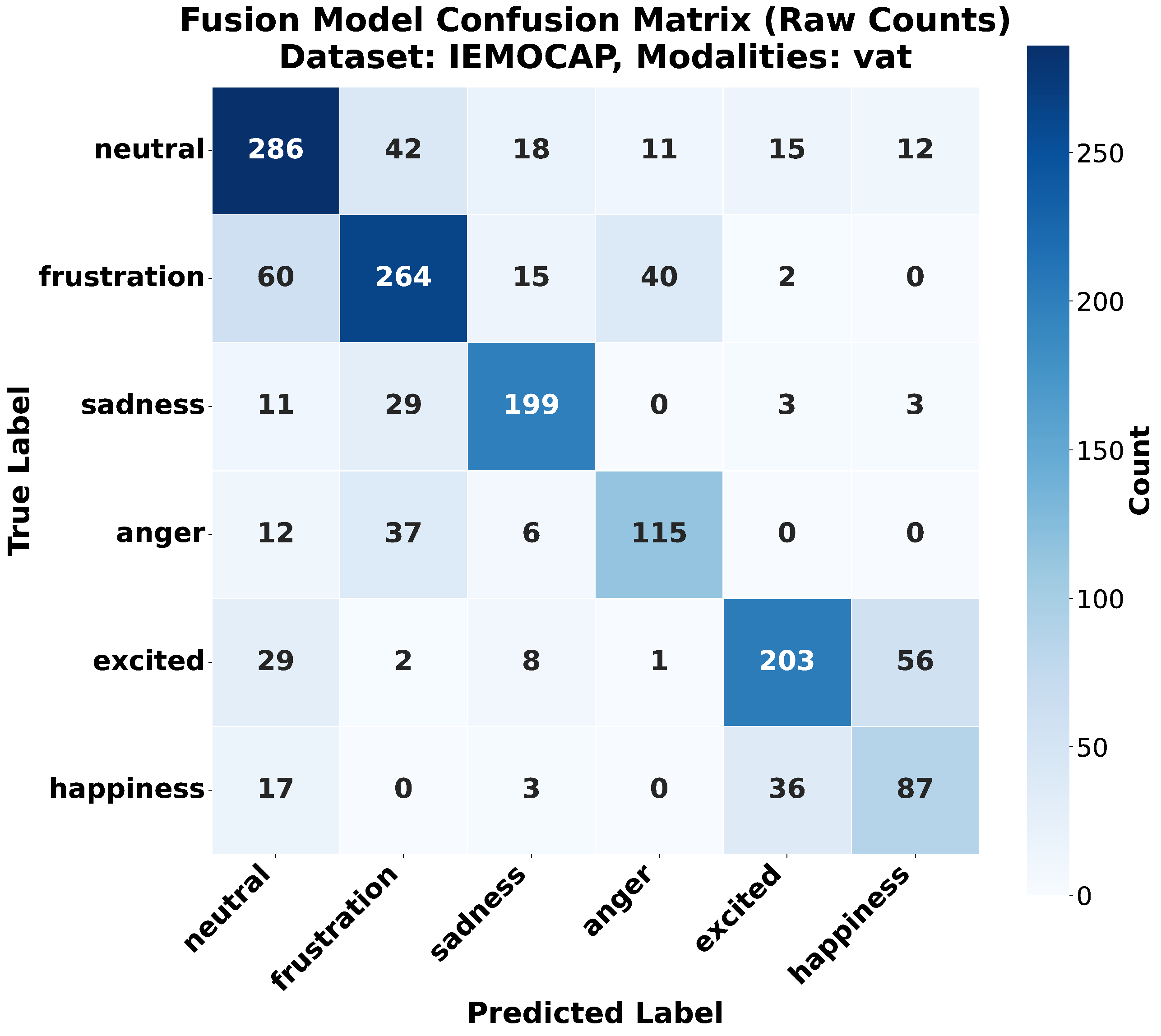}
  \caption{Confusion matrix for emotion classification on IEMOCAP dataset demonstrating balanced performance across emotion categories and clear separation between positive and negative emotional states.}\label{fig:confusion_matrix_iemocap}
\end{figure}

\section{Statistical Significance Analysis}
\label{appendix:stat-test}

We validate our improvement%s 
 on MELD and IEMOCAP through rigorous statistical significance testing against the strongest baselines (TelME and MlutiEMO, respectively) using paired t-tests over 5 independent runs with different random seeds.

\begin{table}[h]
%\small
\centering
\tabcolsep=3pt
\begin{tabular}{llcc}
\toprule
\textbf{Dataset} & \textbf{Method} & \textbf{Mean WF1} & \textbf{p-value} \\
\midrule
MELD & TelME       & $67.30 \pm 0.05$ & \\
     & Our Method  & $\mathbf{67.75 \pm 0.07}$ & \textbf{0.008} \\
\midrule
IEMOCAP & MultiEMO     & $71.18 \pm 0.09$ & \\
        & Our Method   & $\mathbf{72.44 \pm 0.12}$ & \textbf{0.004} \\
\bottomrule
\end{tabular}
\caption{Statistical significance test results. }
\label{tab:significance}
\end{table}

%Results demonstrate statistically significant improvements on both datasets, with particularly strong significance on IEMOCAP (p < 0.01), validating the robustness and consistency of our approach across different experimental runs.

\end{document}